\documentclass[tradiabstract]{aa}
\usepackage{times,amsfonts}
\usepackage{natbib}
\usepackage[english]{babel}
\usepackage[T1]{fontenc}
\usepackage[latin2]{inputenc}
\usepackage{graphicx}
\usepackage{color}
\usepackage{amsmath}
\usepackage{amssymb}
\usepackage{mathrsfs}
\usepackage{aas_macros}

\newcommand\tabcaption{\def\@captype{table}\caption}
\makeatother
\newcommand\gtapprox{\,\lower.6ex\hbox{$\buildrel >\over \sim$} \, }

\newcommand\Nsim{N_{\mbox{\rm \small sim}}}
\newcommand\Npix{N_{\mbox{\rm \small pix}}}

\newcommand{\lbtmp}[1]{(#1)}
\newcommand{\lbdod}[6]{$(l^\circ,b^\circ)=\{\lbtmp{#1},\lbtmp{#2},\lbtmp{#3},\lbtmp{#4},\lbtmp{#5},\lbtmp{#6}\}$ }
\newcommand{\thedodec}{\lbdod{252,65}{51,51}{144,38}{207,10}{271,3}{332,25}}

\newcommand\Sfid{S^{\mbox{\rm \tiny fid}}} %%% EDITOR modify as desired 
\newcommand\Smax{S^{\mbox{\rm \small max}}} %%% EDITOR modify as desired 
\newcommand\nside{n_{\mbox{\rm \small s}}} %%% EDITOR modify as desired 

\newcommand{\twopt}{2-point}

\newcommand\ltapprox{\,\lower.6ex\hbox{$\buildrel <\over \sim$} \, }

\newcommand{\langed}[1]{\textcolor{black}{#1}}
\usepackage{hyperref}
\usepackage{url} 

\providecommand{\url}[1]{\href{#1}{#1}}

\providecommand{\adsurl}[1]{}

\title{A test of the Poincar\'e dodecahedral space topology hypothesis with the WMAP CMB data}
\author{Bartosz Lew \inst{1,2}, Boudewijn Roukema \inst{1}}
\institute{
Toru\'n Centre for Astronomy, Nicolaus Copernicus University, ul. Gagarina 11, 87-100 Toru\'n, POLAND
\and
Department of Physics and Astrophysics, Nagoya University, Nagoya 464-8602, Japan}
\date{Received 2 October 2007 / Accepted 7 January 2008}

\abstract{It has been suggested by Roukema and coworkers (hereafter R04) that the topology of the Universe as probed by the ``matched circles'' method
using the first year release of the WMAP CMB data, might be that of the Poincar\'e dodecahedral space (PDS) model.
An excess in the correlation of the ``identified circles'' was reported by R04, for circles of angular radius of $\sim 11^\circ$ 
for a relative phase twist $-36^\circ$, 
hinting that this \langed{could be} due to a Clifford translation, if the hypothesized model \langed{were} true.
R04 did not however specify the statistical significance of the correlation signal.

We investigate the statistical significance of the signal using Monte Carlo CMB simulations 
in a simply connected Universe, 
and present an updated analysis using
the three-year WMAP data. We find that our analyses of the first and three year WMAP data provide results that are consistent 
with the simply connected space at a confidence level as low as 68\%.
}

\authorrunning{B. Lew}
\titlerunning{Test of Poincar\'e dodecahedral space topology hypothesis.}

\begin{document}
\maketitle
\section{Introduction}

\langed{If the topology of the Universe were multiply connected, as opposed to simply-connected, and if the 
comoving size of the fundamental domain (FD) were smaller than the comoving distance to the surface-of-last-scattering (SLS), 
then it should be possible to detect repeating patterns in the CMB fluctuations 
using full-sky data of sufficient signal-to-noise ratio.}
These fluctuations would be those lying along pairs of circles defined by points of intersection
between different copies of the SLS in the covering space
\citep{Corn98b}.
These patterns although found in different directions of the sky, would constitute so-called 
``matched circles'', as they would represent the same physical points, but observed from different directions due to topological 
lensing.\\
While this principle is true 
for any 3-manifold model of space, 
the number of pairs of ``matched circles'' or
their sizes and relative spatial orientations, as well as their handedness, or phase shift, depend significantly on the assumed
3-manifold and its topological properties, thus 
providing a way 
to observationally distinguish between models.

\par While the positive correlation signal from matched pairs is expected directly from the metric perturbations,
via the Sachs-Wolfe Effect \citep{SW67},
there are many other cosmological effects
(e.g. the Doppler effect, the Integrated Sachs-Wolfe Effect (ISW)) \citep{2008PhRvD..77b3525K},
astrophysical foregrounds \citep{WMAPforegrounds} and instrumental effects that constitute noise, from the point-of-view of a matched circles search,
and the magnitude of the effects depends on the angular scale.
\par
Although the CMB data have been analyzed to detect topological lensing signals since the availability of the COBE 
data \citep{Roukema00-3},
the release of the WMAP observations has provided full-sky data of unprecedented accuracy and resolution,
\langed{opening up more promise for direct tests of the topology of the Universe.  }
Although the ``matched circles'' test is straightforward, it is limited 
due to noise and FD size constraints.
Additional theoretical predictions can be used as independent tests that 
involve predictions of CMB temperature and polarization fluctuations 
\langed{for the case that the Universe is multiply connected},
both in real and spherical harmonic spaces, or
topological effects on the CMB power spectrum \citep{2004CQGra..21.4901A,2003MPLA...18.2099W,2003PhLA..311..319G,2005MNRAS.358.1285D,2004PhRvD..69j3514R,
LLU98,Inoue99,
2007PhRvL..99h1302N,Corn98a,deOliv95,2006PhRvD..73b3511K,2003Natur.425..593L,
2006AIPC..848..774N,2006ApJ...645..820P,
2007AA...476..691C}.
Although a successful ``matched circles'' test would provide strong support for \langed{the Universe being multiply connected}, no statistically-significant evidence
has been found \citep{2004PhRvL..92t1302C,2006astro.ph..4616S}. 

\par
\begin{figure}[!hbt]
\renewcommand{\figurename}{Fig}
\includegraphics[width=0.5\textwidth]{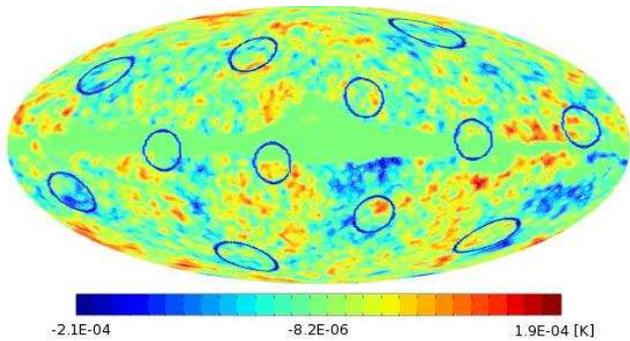}
\caption{Visualization of the matched circles solution reported in \protect\cite{RLCMB04} and reproduced in this paper to
constrain its statistical significance, over-plotted on the first year ILC map masked with Kp2 sky mask.} 
\label{fig:thedodec}
\end{figure}

 In \cite{RLCMB04} we performed a ``matched circles'' search using the first year WMAP ILC map \citep{WMAPforegrounds}
and found an excess correlation, which one would expect under the PDS hypothesis for circles of angular radii $\alpha\sim 11^\circ$
with centers towards \thedodec and their opposites (Fig.~\ref{fig:thedodec}). 

\par In this present work we have two key objectives. We revisit those results,
verify the existence of the excess correlations and quantify their statistical significance. Secondly, we update the search with the WMAP
three year data release, extend it to probe three different resolutions (smoothing lengths) and define the detection confidence thresholds. 
We also discuss the effects of underlying 2-point correlations, smoothing length and incomplete sky coverage on the value of correlation coefficient.

In section~\ref{sec:data_and_sims} we introduce the datasets used in the analysis, 
provide details of their preprocessing, and describe simulations that we use to complete a statistical significance analysis.
In section~\ref{sec:statistics} we introduce the details of the statistics being performed and our confidence-level
analysis. Results are presented in section~\ref{sec:results}. We conclude in section~\ref{sec:conclusions}.

\section{Data and simulations}
\label{sec:data_and_sims}
We perform a ``matched circles'' search using two sets of data. Firstly, for the sake of compatibility, we choose the same
data as in \cite{RLCMB04} -- i.e. the first year WMAP ILC map. 

\langed{The topologically interesting signal generally dominates over the Doppler (and other) components on large scales
\citep{2004PhRvD..69j3514R}: this is a motivation for using a large smoothing length.  }
However, extended flat fluctuations that happen to
have a similar large scale trend can lead to
false positives on large scales \citep{2006astro.ph..4616S}.  This
implies a trade off in the choice of the smoothing length of the data,
between large smoothing lengths preferred by the topologically
interesting content, and small smoothing scales which avoid false
positives induced by chance correlations of extended flat fluctuations.

We choose to test three different smoothing scales: $FWHM\equiv\lambda\in\{1^\circ,2^\circ,4^\circ\}$. 

The ILC map was obtained from \langed{a linear combination of 
one degree smoothed maps in the five frequency bands, by inverse noise co-adding them,}
hence its resolution is consistent with a one degree smoothing scale.
\langed{
We further Gaussian smooth this map in spherical harmonic space by convolving it with Gaussian beam response kernels of FWHM corresponding to
$2^\circ$ and $4^\circ$ respectively, to obtain the first set of data for matched circles tests.}

\par \langed{Secondly}, we choose the three year foreground reduced WMAP data from individual frequency bands Q[1/2],V[1/2] and W[1/2/3/4] and 
co-add them into one map, according to the inverse noise weighting scheme \langed{used in} \cite{2003ApJS..148..135H}. 
We call the resultant \langed{map the ``INC map''.}
\langed{We smooth the INC map using a Gaussian convolution kernel to four different FWHM smoothing 
lengths: $\lambda\in\{0.5^\circ,1^\circ,2^\circ,4^\circ\}$.}

\par We downgrade all data from the initial resolution defined by the Healpix pixelization scheme \citep{Healpix2005} with resolution 
parameter $n_s=512$  (res. 9) to a resolution parameter of $n_s=256$ (res. 8).
We remove the the residual monopole and dipole components ($\ell=0,1$) in spherical harmonic space\footnote{Since 
the residual WMAP maps foregrounds are strong, we perform this step using the Kp2 sky mask to keep 
the compatibility between the data and simulations.}, 
because these components are of no cosmological interest.
At the final stage of preprocessing, we remove the residual monopole by offsetting the maps in real 
\langed{(2-sphere)} space
so that $\langle T\rangle = 0$ outside the Kp2 sky mask.

\par Throughout the analysis, \langed{i.e. for both the ILC and INC maps}, we use the Kp2 sky mask, which masks $\sim 15$\% of the sky including the brightest \langed{resolved} point 
sources.
\langed{The Kp2 sky mask
is} different from the sky mask used in ~\cite{RLCMB04}.
\langed{While the ILC map is best suited e.g. for the full sky low multipoles alignment analysis, for the purpose of the matched 
circles test, 
the residual galactic contamination should be masked out, although we realize that the use of the Kp2 mask \langed{may} be too conservative.}
\langed{In App.~\ref{sec:S-gal-cut} 
we compare the impact of different sky masks and 
demonstrate that our results are not very sensitive to the precise characteristics of the sky mask.}

\langed{
For each of the two data sets, we produce $\Nsim=100$ realistic Gaussian random field (GRF) signal and noise simulations of the WMAP data
to quantify the statistical significance of plausible detections, to discard false positives, and to resolve 
the $2\sigma$-confidence levels.}
\langed{
Therefore, for the first dataset we simulate the first year ILC map, inside ``region 0'' defined outside the Kp2 sky mask of \cite{WMAPforegrounds},
and for the second dataset we simulate the three year WMAP INC map.
}
\par As will be shown in Sect.~\ref{sec:statistics}, the matched circles correlation coefficient depends on 
the monopole value in the map. Also, in principle it is sensitive to the shape of the \twopt correlation function, 
since the correlator is a \twopt statistic, by construction, and so it becomes a measure of the underlying intrinsic
\twopt correlations in the CMB
(albeit via a specially selected subset of pairs of points on the matched circles). 
Therefore it is necessary to take into account possible variations in the underlying \twopt
correlation function with varying angular separation, which if not properly accounted for in simulations may lead to 
under(over)-estimation of the confidence level thresholds.

\langed{Given that the concordance best fit LCDM cosmological model \citep{2007ApJS..170..377S} yields a very poor fit to 
the CMB data at large angular scales, due to lack of correlations in the \twopt correlation function of the data with respect to the LCDM model 
at scales $>60^\circ$, and
that the correlation statistic is sensitive to the details of the intrinsic \twopt CMB correlations 
(and in particular to any large scale anomalies),
we do not assume the LCDM model to help create our simulations. Instead, we take a model independent approach.}
\langed{As the CMB reference power spectrum in our GRF simulations of the expected signal,
we use the reconstructed power spectrum from the three-year WMAP data \citep{2007ApJS..170..288H}}
\footnote{In the high $l$ end (noise dominated range) of the reconstructed power spectrum, the unphysical negative values
are zeroed to have a zero contribution to the total variance of the map. This approximation has a negligible 
effect due to small statistical weight of the large l multipoles, and large exponential Gaussian smoothing that we apply to the data.
\langed{In practice, this approximation has a negligible effect on the variance of the resulting simulation. 
Moreover, it can at most} only make
our analysis more conservative. 
}.
\langed{Furthermore, we neglect the effects of cosmic variance, and only randomize the phases (and noise realizations)
in our simulations. We remove the $C_{\ell=0,1}$ (i.e. the monopole and dipole) components from our simulations.\\}
We use the same set \langed{of $a_{lm}$s representing the CMB signal} for a single simulation of the two datasets, followed by convolution with instrumental beam profiles.\\
\langed{For each differential assembly (DA), we simulate the noise according its properties  and scanning 
strategy (number of observations per pixel in map) using uncorrelated, Gaussian noise.}

\par The simulations are preprocessed in exactly the same way as the observational data. 
\langed{We neglect the impact of the (resolved or unresolved) 
point sources which is negligible, since we apply relatively large smoothing and use the Kp2 sky mask for the analysis.}

In Appendix A, we discuss the sensitivity of our results to the degree of smoothing, 
the sky mask applied, and the assumed statistical approach in greater \langed{detail}.

\section{Statistics}
\label{sec:statistics}
We describe our correlator statistics, parameter space, search optimization and approach for assessing the statistical significance.

\subsection{Matched circles test}
As in \cite{2004PhRvL..92t1302C} and \cite{RLCMB04} we use a correlation \langed{statistic} of the form
\begin{equation}
S = 2 \frac{\langle T_i m_i T_j m_j\rangle}{\langle T^2_i m_i m_j\rangle+ \langle T^2_j m_i m_j\rangle}
\label{eq:Sstat}
\end{equation}
where the index $i$ defines a set of all points in the ``first'' set of six circles related to the orientation of a fundamental 
dodecahedron; index $j$ is the set of corresponding points along the matched six circles; 
and $m_{i}, m_j$ \langed{are cut} sky weights of the Kp2 sky mask, which can have a value of either $0$ for a masked pixel or
$1$ for an unmasked pixel. 
\langed{Clearly, perfectly matched circles would yield $S=1$, which, due to non-zero  noise contributions, is %, of course, 
not possible in reality.}

\langed{
The dispersion of the correlation coefficient as defined in Eq.~\ref{eq:Sstat} 
is statistically enhanced in the small circles regime, 
due to the joint effect of 
the reduced number of points probing the matched circles as compared to larger circles,
the accidental correlations of large (w.r.t. the smoothing scale) flat fluctuations
that happen to have similar (or opposite) large scale trends, as well as 
due to the fact that the r.m.s. values necessarily shrink (down to zero in case of zero mean fluctuations) 
for circles of size comparable or smaller than the smoothing length. 
}
As shown in Sect.~\ref{sec:results}, this
reduces the ability to robustly determine the \langed{degree} of consistency or inconsistency of data with simulations,
due to the finite accuracy of the $S$ values and significant steepening of confidence-level contours in this regime.

We note that the $S$ statistic value would tend to unity, regardless of the shape of the 
underlying CMB fluctuations, as the monopole increases in the CMB maps. One could expect \langed{a} similar effect
for \langed{the} dipole component, for small circles. 
This effect would affect the simulations and the data to the same
extent. The sensitivity of the test would however be significantly weakened and as such we remove the monopole and dipole
components from the datasets for the analysis, and defer study of the impact of other small $\ell$ multipoles
to appendix~\ref{app:Sdependences}.

\subsection{Parameter space}
\label{sec:parspace}
We perform a resolution-limited, full parameter-space search over the orientation of the fundamental dodecahedron, and 
over a limited range of the identified circles sizes of up to $20^\circ$.
The parameters are defined as follows: $l,b$ -- galactic longitude and latitude of the first circle, 
$g$ -- the angle of rotation of the dodecahedron about the axis determined by $(l,b)$, 
$a$ -- the angular radius of the matched circle, and 
$s$ -- the twist parameter defining the relative phase offset of the matched circles. \\
We use the following parameter space:
\begin{equation}
\begin{array}{ccl}
l &\in& [ 0^\circ,72^\circ)\\
b &\in& [ ~26.57^\circ, 90^\circ)\\
g &\in& [ 0^\circ,72^\circ)\\
a &\in& [ 1^\circ,20^\circ]\\
s &\in& \{-36^\circ, 0^\circ, 36^\circ\}\\
\end{array}
\label{eq:param_grid}
\end{equation}

The boundaries in $(l,b)$ \langed{conservatively} cover a larger region than the
one twelfth of the sphere from which a ``first'' circle centre can be chosen
non-redundantly. The range of angle $g$ is 72$^\circ$,
to cover all possible orientations of the fundamental dodecahedron for a chosen ``first'' circle centre.
Values larger than $72^\circ$ would yield the same set of 12 circle centres as a rotation by that angle modulo $72^\circ$. 
The interval in circle size $a$ is chosen to be roughly symmetric and centered about the $11^\circ$
value suggested by \citet{RLCMB04}. The three twists $s$ are chosen as in \citet{RLCMB04}.

For all datasets we use the same resolution of $1^\circ$ in probing the parameter space, except for the
data with smoothing length $\lambda=0.5^\circ$, in which case we use a resolution of $0.5^\circ$.

\subsection{Accuracy and search optimization}
\begin{figure*}[!hbt]
\centering
\renewcommand{\figurename}{Fig}
\includegraphics[angle=-90,width=0.49\textwidth]{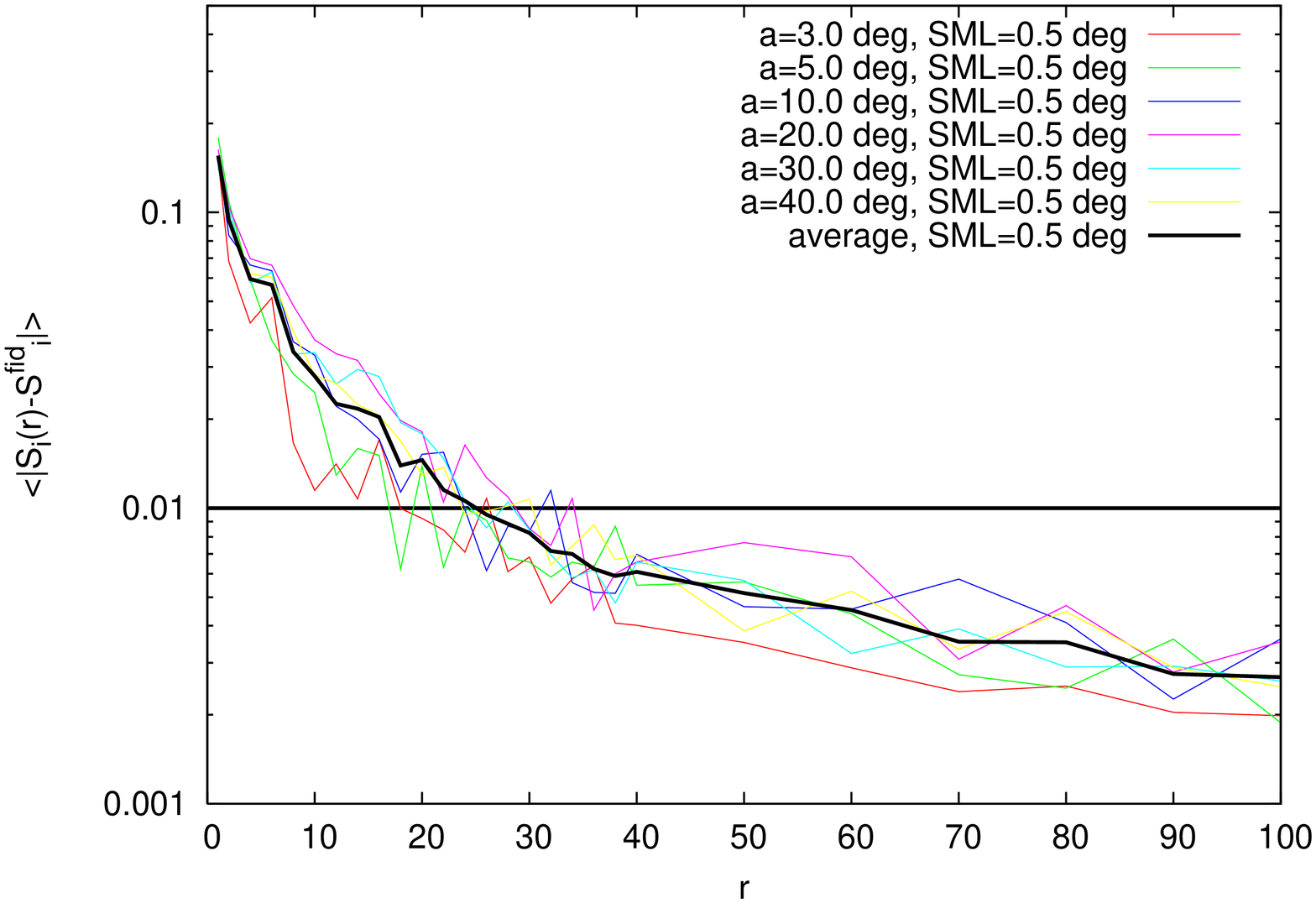}
\includegraphics[angle=-90,width=0.49\textwidth]{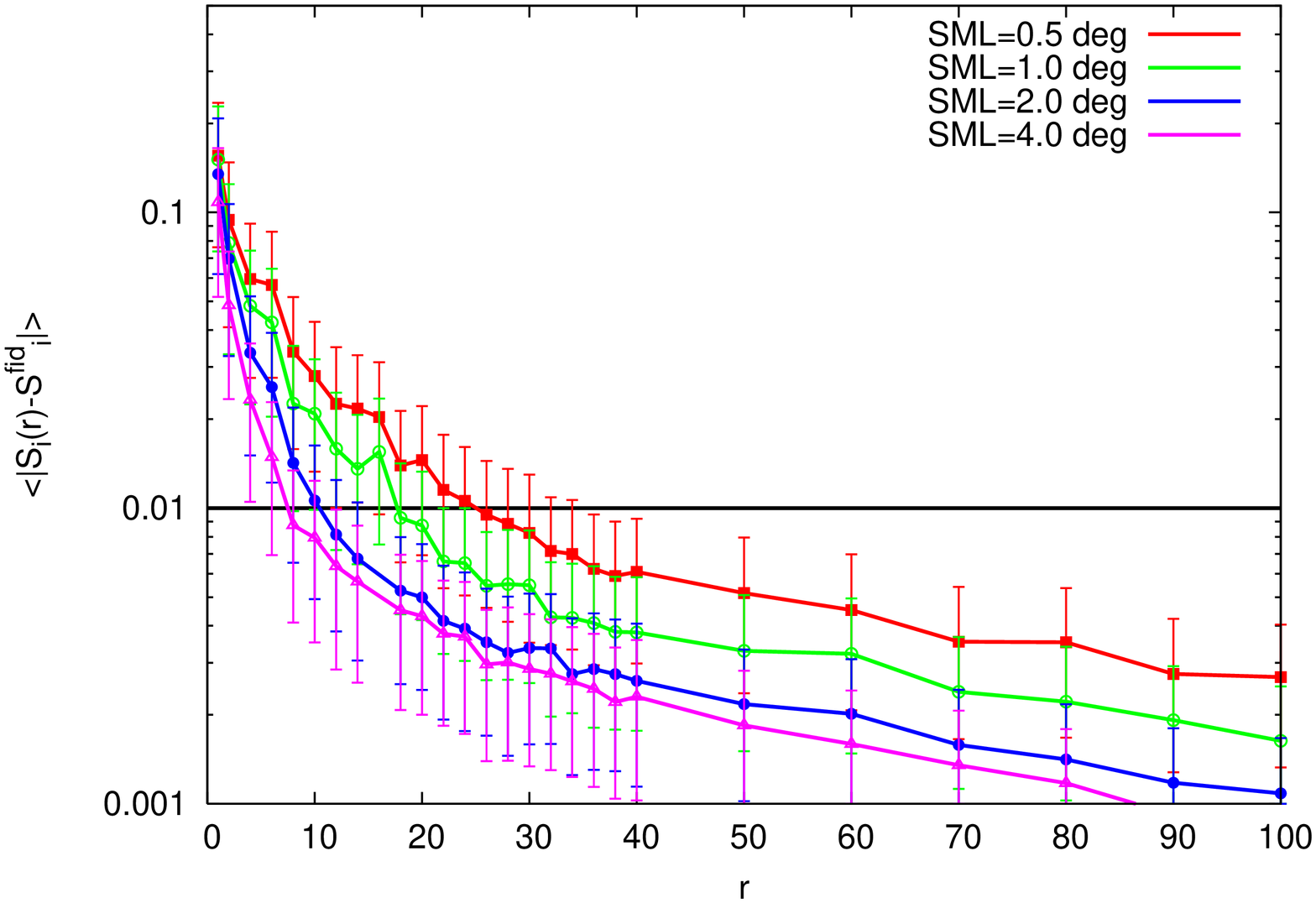}
\caption{Left panel: Convergence of $S$ values to the ``ideal'' fiducial value $\Sfid =S(r=1000)$  as a function of resolution parameter $r$ 
(a sampling density resolution parameter, 
defining the number of pixels to be used, to probe the CMB fluctuations along circles through Eq.~(\protect\ref{eq:Npix})) and as a function of circle size $a$. 
The $\Npix(a)$ function shape for a 
given smoothing length is fitted linearly, so that the accuracy in $S$ values was approximately constant for all considered circles radii.
The assumed working precision level of $\Delta S=0.01$ is marked with thick horizontal line. 
For clarity, only $\Delta S$ relation derived for data smoothed with Gaussian $\lambda=0.5^\circ$ is shown. 
Similar relations are obtained for the remaining three smoothing lengths.
The average value (black thick line) from all tested circles radii is used to define the required value of $r$ parameter
for the circle search with data smoothed to $0.5^\circ$, in order to achieve the targeted accuracy on $S$ value.
Right panel: Average $S$ convergence relations derived for data with different smoothing lengths $\lambda\in\{0.5^\circ,1^\circ,2^\circ,4^\circ\}$, along
with $1\sigma$ error bars from 20 simulations. 
The intersections of these with the horizontal (black thick) line give the required values of $r$ for each smoothing length in order to obtain the
assumed working precision of $\Delta S=0.01$.
} 
\label{fig:convergence}
\end{figure*}
The resolution of the data that we analyze \langed{is spatially constant and} is limited by the finite pixel size, 
so circles of different sizes are probed by different \langed{numbers} of pixels. \\
\langed{As the parameter space of the search is large, it is important to consider the trade-off between the accuracy of the estimates of $S$
(directly related to the number of pixels probing the underlying fluctuations) and the numerical computational time needed to obtain
better accuracy. However, the speed of the search can be substantially increased, since the effective 
resolution of the data in our case is not limited by the pixel size, but rather by the smoothing length.}

\par
In this section, we focus on the density of points (probing the fluctuations along circles in the sky) required to obtain
a given accuracy in estimating $S$, and its dependence on \langed{the angular radius of the circles $a$,} a circle sampling density parameter $r$, 
the map resolution parameter $\nside$, and the smoothing length properties of the data.

\langed{Maps smoothed with larger smoothing lengths have fewer significant, high-spatial frequency Fourier modes, 
and there is no need for fine sampling in order to fully encode the information content along the circles.  }
Assessing the same level of 
precision for smaller circles also requires a smaller number of pixels than for larger circles.  
 
We \langed{perform} a series of tests to determine the sampling density required to achieve our desired accuracy level.
The tests rely on measuring the speed of convergence to the ``ideal'' fiducial $\Sfid$ value, 
derived using far more points in the circle than the number of available pixels along the circle in our datasets\footnote{
For 
all directions pointing inside a single pixel, the same temperature value of that pixel is used.}, as a function of the increasing sampling density. 

\langed{We empirically model the circle sampling-density function} in such a way that
\langed{for a given} $r$ value parameter, and for a given smoothing length of the data,
 the accuracy of the resulting $S$ values (i.e. the statistical size of the departure from the fiducial value) 
is approximately the same for all circle sizes (Fig.~\ref{fig:convergence} left panel).
We use the following fitted function:
\begin{equation}
\Npix = \Bigl(3.40 a [deg] + 76.85\Bigr) \Bigl(\frac{r}{32}\Bigr)\Bigl(\frac{n_s}{256} \Bigr)
\label{eq:Npix}
\end{equation}
where $\Npix$ is the number of pixels used for calculation of $S$, for a circle of angular size $a$, and for a map of resolution
$\nside$. The resolution parameter $r$ controls the sampling density.
In practice, we choose the closest, even integer as an $\Npix$ value for the calculations.
This empirically-devised formula yields approximately the same accuracy of derived values of S for all circle sizes
(Fig.~\ref{fig:convergence} left panel), and holds for all smoothing lengths.
The aim is to find a value of $r$, for each smoothing length, which will provide 
sufficient accuracy.

\par We therefore calculate the deviation $\Delta S(r)$
\begin{equation}
\Delta S(r) = \langle|S_i(r)-\Sfid_i(r=1000)|\rangle
\label{eq:convergence}
\end{equation}
where the $\langle\rangle$ averaging is performed over all curves derived from Eq.~\ref{eq:Npix} for circle radii
$a [deg] \in\{3,5,10,20,30,40\}$. 
\par We assume the working accuracy for $S$ values to be $\Delta S=0.01$ throughout the analysis.
This defines the required values of \langed{the} sampling density parameter $r$ (Fig.~\ref{fig:convergence} right panel) and 
the corresponding number of pixels to be used (Eq.~\ref{eq:Npix}) to achieve the targeted accuracy. 
For the smoothing lengths $\lambda [deg] \in\{0.5,1,2,4\}$, the required resolution parameter values are
$r\in\{26,18,12,8\}$. We use these values throughout the analysis with both the data and the simulations.

\subsection{Statistical significance}
In this section we discuss our statistical approach for quantifying the confidence intervals.

\par Since our simulations simulate CMB fluctuations in an isotropic, simply-connected Universe, we 
test the consistency of the WMAP data with the null hypothesis that the CMB is \langed{an arbitrary} realization of the 
GRF in a simply-connected space. We quantify the degree of consistency via $S$ correlator values
obtained from the data, and compared with those of simulated distributions from $\Nsim = 100$ GRF simulations (Sect.~\ref{sec:data_and_sims}).
As an alternate hypothesis we choose the PDS topological model. The inconsistency of the data with the 
simulations, at high significance level, would then be considered as consistency in favor of the alternative hypothesis (PDS model).

\par Since we are interested only in the highest positive $S$ correlations, 
we build probability distribution functions (PDFs) of $\Smax(a)$, the maximal value of the correlation $S(a)$ found in the 
 matched circle search in the parameter space $(l,b,g,s)$ (Eq.~\ref{eq:param_grid}), using $\Nsim = 100$ simulations\footnote{
Statistically there are some small differences in the $S$ values resulting
from probing slightly different angular separations (arising due to
different separations of pairs of points, for the same pair of matched
circles, when calculated with two different phase twists: $s=0$ and $=\pm
36$), due to the dependence of the $S$ value on the underlying CMB \twopt{}
correlation function. For such small twists, this is found to be
of the same magnitude as the statistical error
on the $\Delta S(\approx 0.01)$ for all considered circle radii.}.
We probe the underlying PDFs of $\Smax(a)$ at $8$ different values of $a$, i.e. for $a \in \{1,2,5,8,11,14,17,20\}$ \langed{in degrees.}

We reconstruct the confidence intervals $[c(a),d(a)]$, for the 68\% and 95\%  confidence levels
defined by the (cumulative) probability $P$ of finding a GRF, simulated, CMB realization
that yields $\Smax_{\mbox{\rm \tiny sim}} > \Smax_{\mbox{\rm \tiny data}}$:
\begin{equation}
\begin{array}{ccc} 
P(\Smax_{\mbox{\rm \tiny sim}} > \Smax_{\mbox{\rm \tiny data}})(a) &=& 1 - \int\limits_{c(a)}^{d(a)} f(\Smax,a) d\Smax\\
&=& 1 - \sum\limits_{i=1, \Smax_{\mbox{\rm \tiny sim,i}} \leq \Smax_{\mbox{\rm \tiny data}}}^{i=\Nsim} 1/\Nsim
\end{array}
\end{equation}
where $c(a) = \mbox{min}(\Smax)(a)$ and $f(\Smax,a)$ is the MC probed PDF of the $\Smax$ values.

\par We interpolate confidence interval contours for the remaining $a$ values of the parameter space using
4th order polynomial fit.

\par In the next section we apply this procedure to the considered WMAP datasets and simulations and present our results.

\section{Results}
\label{sec:results}
\begin{figure*}[!hbt]
\renewcommand{\figurename}{Fig}
\includegraphics[angle=-90,width=0.48\textwidth]{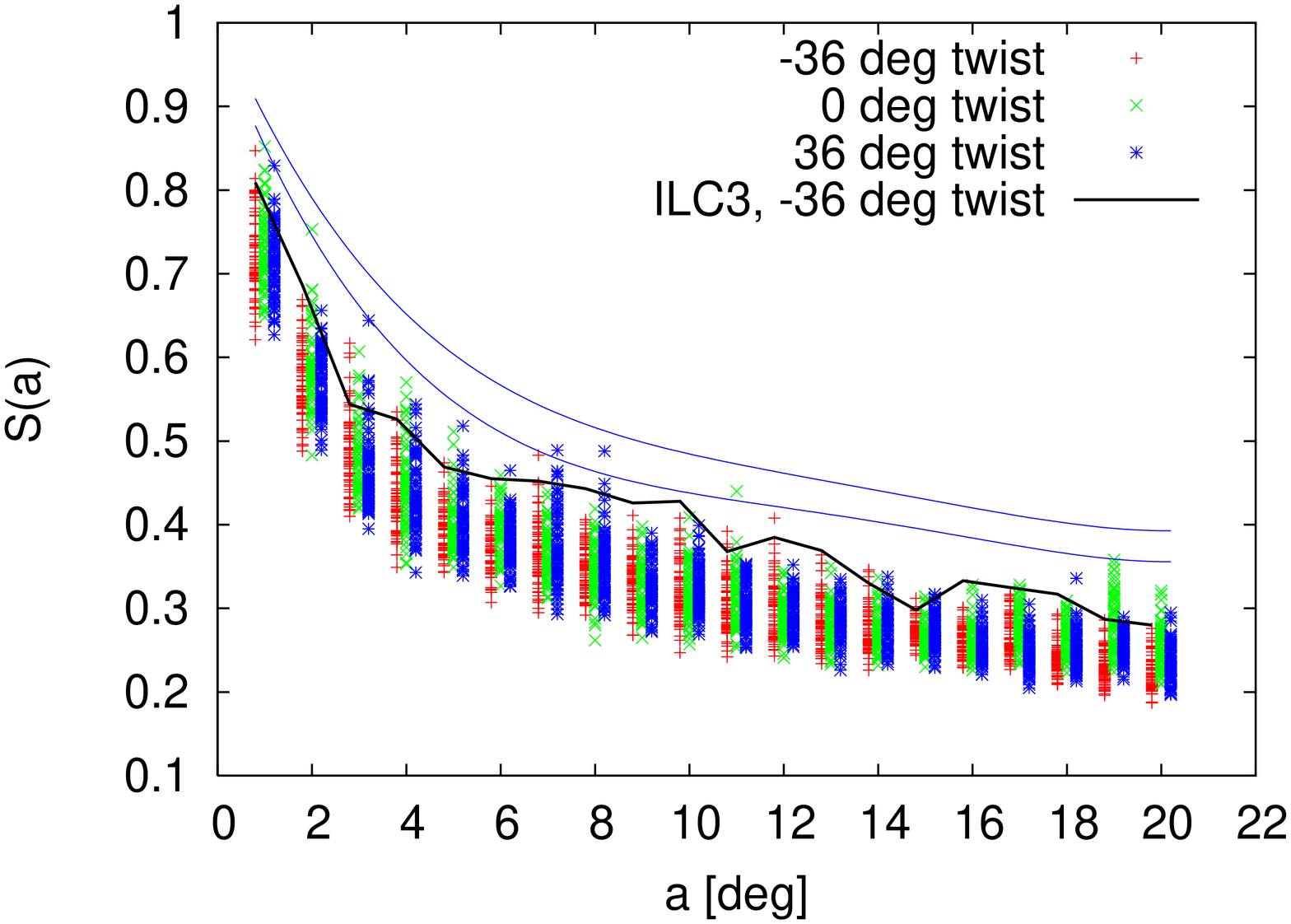}
\includegraphics[angle=-90,width=0.48\textwidth]{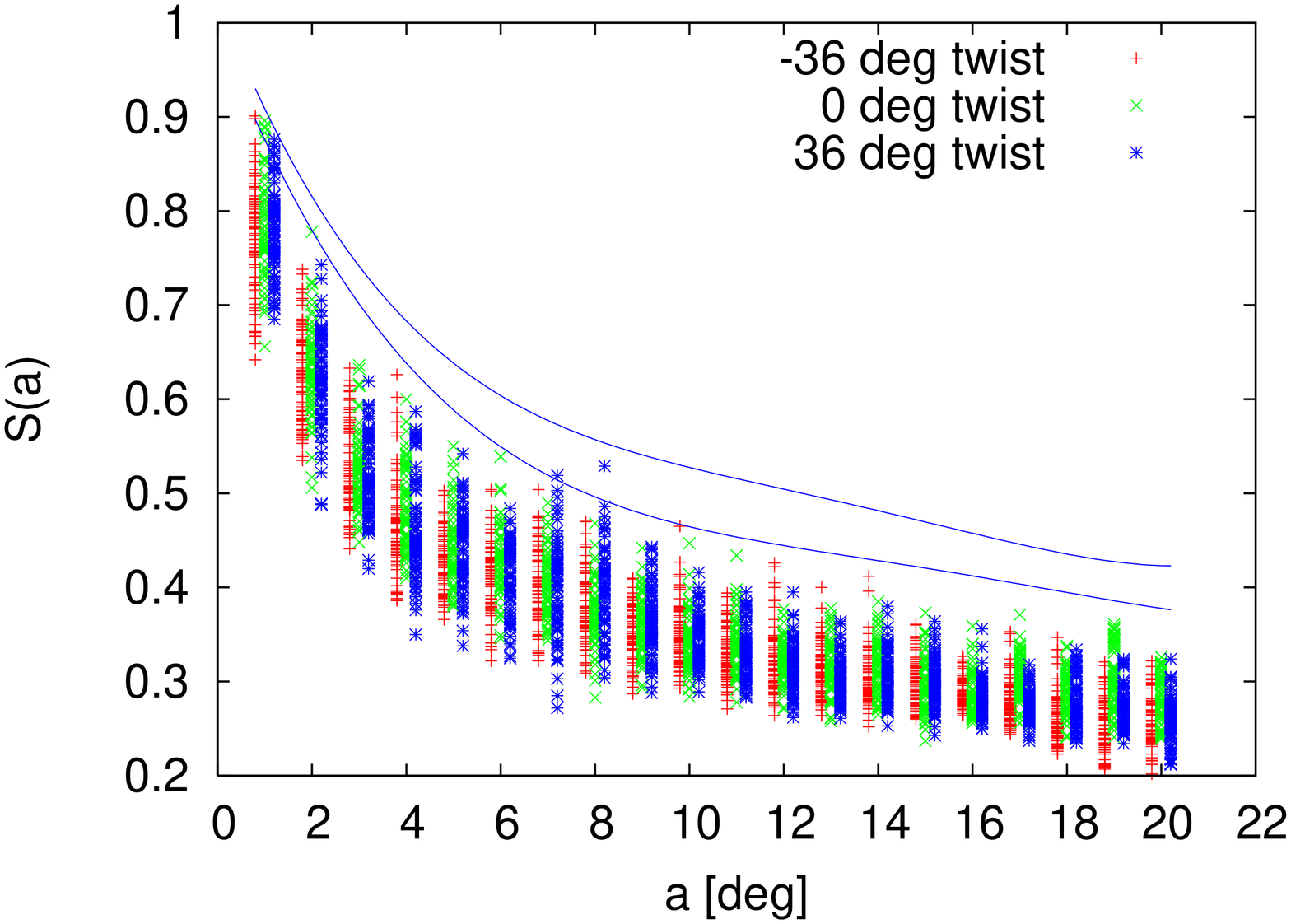}\\
\includegraphics[angle=-90,width=0.48\textwidth]{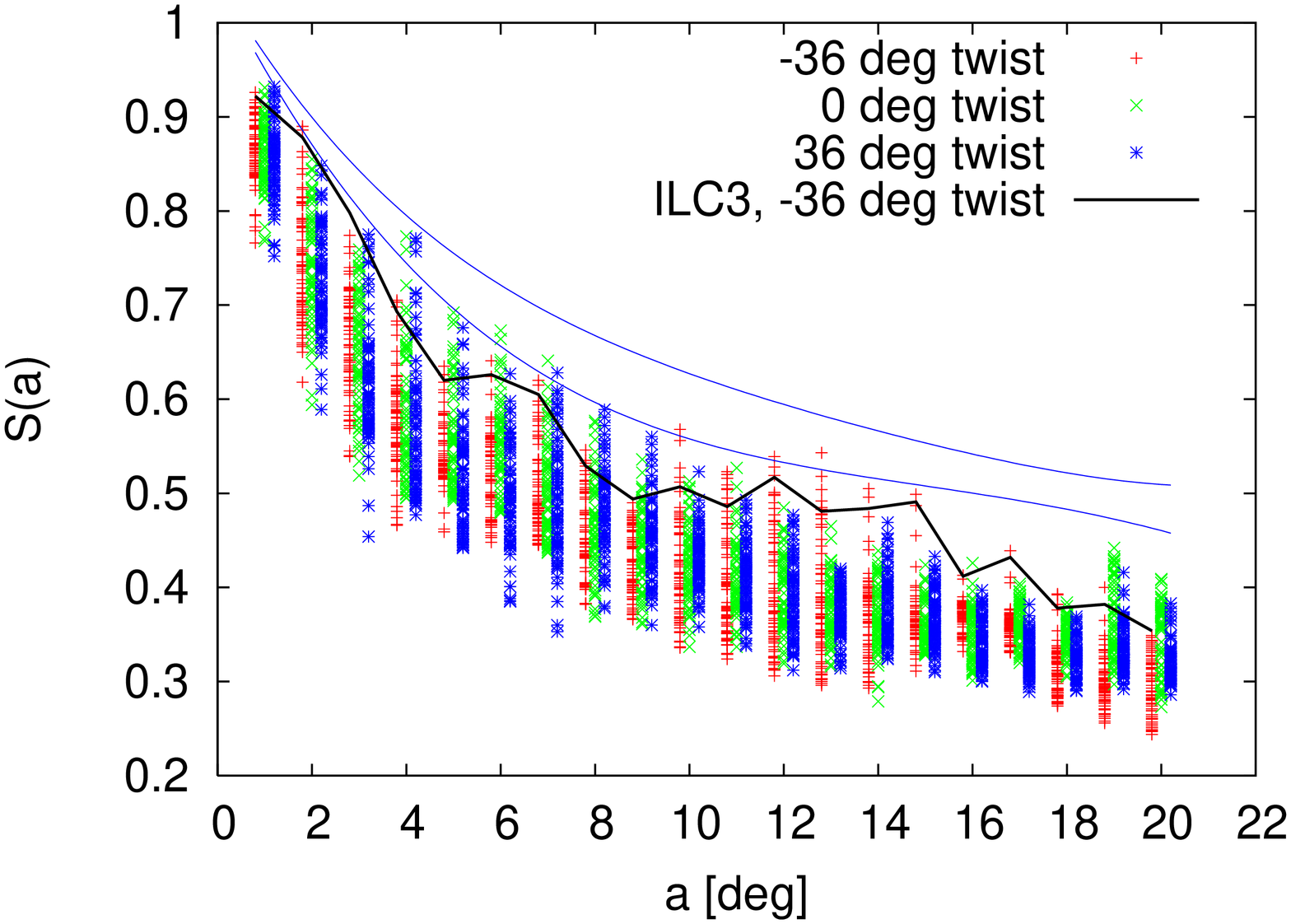}
\includegraphics[angle=-90,width=0.48\textwidth]{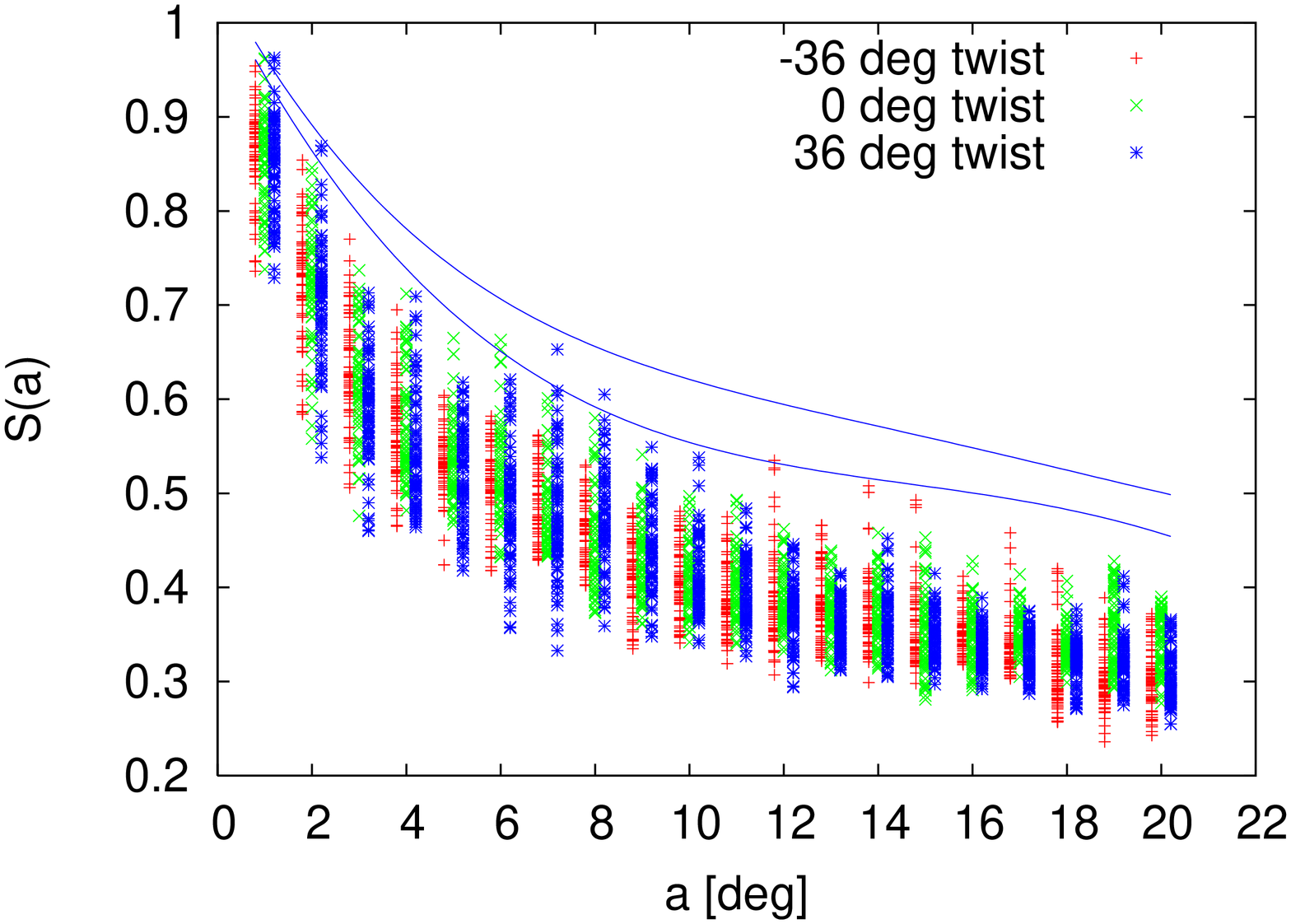}\\
\includegraphics[angle=-90,width=0.48\textwidth]{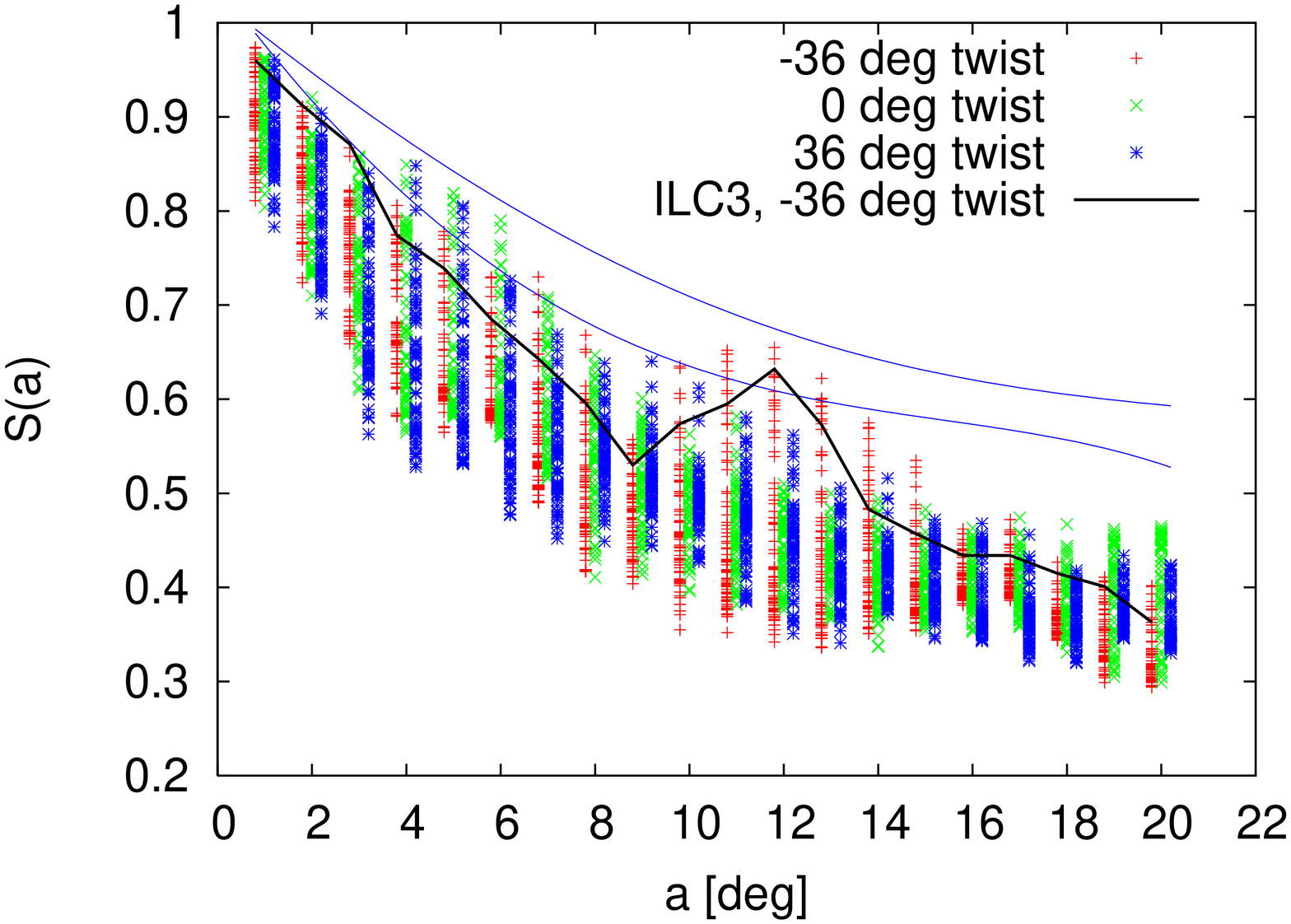}
\includegraphics[angle=-90,width=0.48\textwidth]{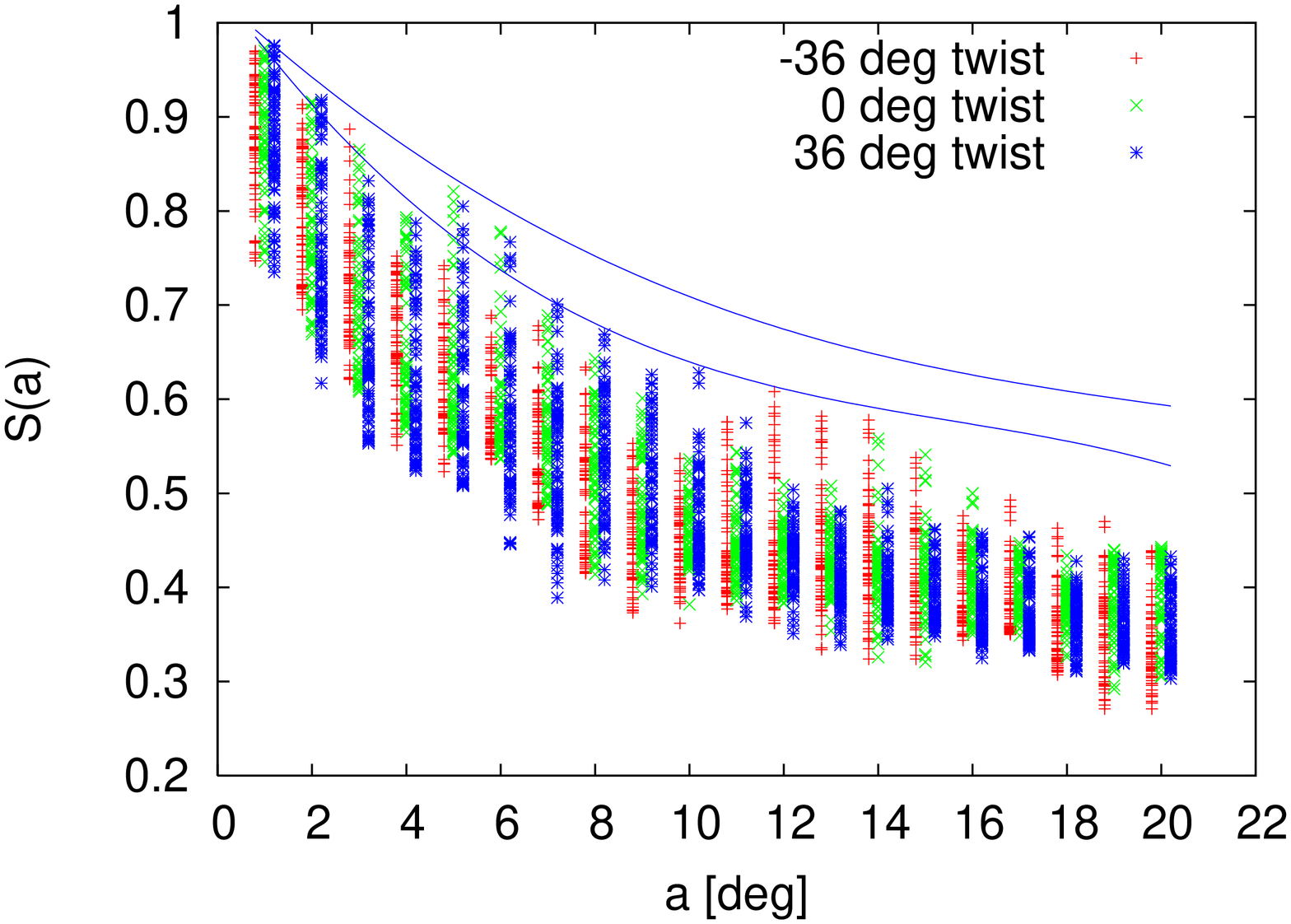}
\caption{Results of the search in the parameter space (see Sect.~\protect\ref{sec:parspace}) for the highest $S$ correlations in the first year WMAP ILC map (left column) 
and three year INC map (right column) smoothed to 
$\lambda =1^\circ$ (top), $\lambda=2^\circ$ (middle), $\lambda=4^\circ$ (bottom). For clarity only the highest 72
$S(a)$ statistic values are plotted for each of the three considered phase shifts: $-36^\circ, 0^\circ, 36^\circ$ marked with 
red ($+$), green ($\times$) and blue ($*$) respectively, and separated by $0.2^\circ$ offset for better visualization and comparison.
The red crosses correspond to the PDS model. 
The thick solid line in the left column show the $\Smax$ values for a search in the three-year ILC data with a phase shift
of $-36^\circ$.
The $68\%$ and $95\%$ confidence level contours from $\Nsim =100$ simulations
are over-plotted. Clearly we reproduce the results of \protect\cite{RLCMB04}. Most of the points with the highest $S$ values in the 
range of $10^\circ\leq a\leq 12^\circ$ closely correspond to the solution depicted in Fig.~\ref{fig:thedodec}. 
It is easily seen \langed{that much} higher correlation coefficients would have been required in order 
to significantly reject the null hypothesis that the Universe is simply connected in favor of the PDS model alternative hypothesis.} 
\label{fig:ILC1INC3}
\end{figure*}

\langed{In Fig.~\ref{fig:ILC1INC3} we present results of the all-parameter-space search for the WMAP first year ILC map (left panel),
and the three year WMAP INC map (right panel).}

\par The signal at $\sim 11^\circ$ in Fig.~4 of \cite{RLCMB04} is reproduced and plotted with red crosses in Fig.~\ref{fig:ILC1INC3}
(middle-left).

\langed{Clearly, it is not necessary to process large number of simulations to resolve high confidence level contours, since all the 
datasets are consistent with the simply connected space GRF simulations at a confidence level as low as about 68\% at all smoothing scales.}

\par Is is easily seen that as the circle size shrinks to zero ($a\leq 2^\circ$), 
it is difficult to estimate precisely the significance of the detections since the CL contours steepen, 
while the accuracy of the S value determination is fixed at $\Delta S \sim 0.01$. This effect is most severe for large
smoothing scales, as expected.

\par We note that the correlations $S$ tend to increase in relation to the smoothing length applied to the data. \\
In particular, the signal reported in \cite{RLCMB04} is sensitive to increases in the smoothing length.
While at the smoothing length of $\lambda=1^\circ$ there is practically no excess maximum in $\Smax (a \approx 12^\circ)$ 
for $s=-36^\circ$ relative to 
\langed{the $\Smax$ values for $s=0^\circ$ and $s=+36^\circ$,
on the other hand,}
the excess is clearly seen at the smoothing length of $\lambda=4^\circ$, where
its significance increases almost up to the 95\% CL.

\langed{The results for the three year INC data} are consistent with the first year data,
in \langed{the} sense that no statistically important excess correlations are found.

\langed{In addition to analysing our two primary datasets, 
we also carried out the following complementary searches.}

\langed{We completed an all-parameter-space search using} the three year WMAP ILC data. We find that the excess correlation 
corresponding to the hypothesized PDS model is 
weakened for all smoothing lengths (for $s\in\{0^\circ,+36^\circ\}$), and basically 
\langed{indistinguishable from the noise of what would be false positive detections if we were to define the 68\% confidence level as a detection threshold}.
\langed{For $s=-36^\circ$, we} plot the $\Smax(a)$ values for the three year ILC data with \langed{a} black line in the left column
of Fig.~\ref{fig:ILC1INC3}.

\par \langed{Our other complementary test was that} we performed a $0.5^\circ$ resolution all parameter space search, using 
first and three year INC data and did not find extra strong localized correlations. However since the computation time
increases with the power of the increased resolution (i.e. increasing the resolution by a factor of two 
increases the calculation time by a factor of $2^{n}$ where $n=4$ is the number of parameters in parameter space) 
we haven't performed the significance analysis with simulations, and therefore we do not present these results.

\section{Discussion}
\label{sec:discussion}

The analysis of the correlations derived from the data and presented in the previous section finds no statistically-significant 
detections.
The cross-correlations of the $S$ values, obtained for different angular radii of the matched circles,
were however neglected. It is of course faster to compute confidence intervals for a sparse parameter space and interpolate in between. 
However, the significance of any detections found this way \langed{(i.e. conditionally to the \langed{ {\em a priori} } assumed circle radii)}
might be overestimated, compared to the case when all possible correlations were accounted for in the full covariance matrix
analysis. In the present work, since we do not find any significant deviations from the null hypothesis (i.e. we do not find
any strong outliers in the $S$ correlations) in any individually probed value of the ``$a$'' parameter, we 
find no need for any further extensions to the significance analysis already pursued.

\langed{We note that these cross-correlations are present not just in the
data, but also in the Monte-Carlo simulations, so they 
 affect the analyses of simulations and data to the same extent.}

Finally, we note that our statistical approach of considering only the maximal $S$ correlation
values could be altered to consider the full distributions of $S$ correlations. 
\langed{However, we} are especially interested in \langed{viable} candidates for non-trivial 
topology (especially in the proposed correlation signal around angular \langed{circle} radii of $11^\circ$) 
and as such, the models with the largest $S$ values are the best candidates. 
Given that there can be only one correct orientation of the fundamental dodecahedron and hence only one $S$ correlation value
corresponding to it (most likely the largest locally found $S$ value), 
in the alternative way involving the full distributions of $S$ correlations,
the test would be heavily dominated by numerous values that will not be associated 
with the true topological correlation signal. As a result, the test would mostly measure a degree of 
consistency between the simulations and data with respect to underlying two-point correlations via the circles on the sky, 
rather than candidates for non-trivial topology. 
We therefore build and rely on the statistics of specifically 
selected (in the full parameter space search) $S$ values, one for
a given simulation at a given circle radius, 
to build our statistics and reconstruct
the confidence thresholds.
Although we are aware that relying on the distributions of maximal values of random variates may lead to 
asymmetrical  distributions with enhanced tails, we note that in the case of the $S$ statistics, the possible values are by definition 
restricted to within the range $[-1, 1]$. 
We also note that it is unnecessary to resolve high confidence level contours, since the
data are consistent with simulations mostly to within  $\sim 1\sigma$ confidence contour.

\section{Conclusions}
\label{sec:conclusions}
\par In \cite{RLCMB04} it has been suggested that the shape of the space might be consistent with \langed{the}
Poincar\'e dodecahedral space (PDS) model (Fig.~\ref{fig:thedodec}).
This suggestion was due to an \langed{excess positive correlation} in the matched circles test \citep{Corn98b} of \langed{the} first year 
WMAP ILC map, however the statistical significance of this excess was not specified.

We have revisited those results and found consistent correlation excess corresponding to the same orientation of the fundamental dodecahedron using
independent software. 

We extended and updated the matched circles search with the WMAP three year ILC data and the three year foreground reduced, inverse noise co-added map, 
and tested these at three different smoothing scales $(1^\circ,2^\circ,4^\circ)$.

We performed \langed{an} analysis of the statistical significance of the reported excess, based on
 realistic and very conservative MC GRF CMB simulations of the datasets.

We find that \langed{under} ``matched circles'' tests, both the first and three year WMAP data are consistent with the 
simply-connected topology hypothesis, for all smoothing scales, at a confidence level as low as 68\%, 
apart from the first year ILC data smoothed to $4^\circ$, which \langed{are} consistent at 95\% CL.

\begin{acknowledgements}
\langed{The authors} would like to thank the anonymous referee for useful comments and suggestions.
BL would like to thank Naoshi Sugiyama for his support. 
BL acknowledges the use of the computing facilities of the Nagoya University (Japan), and support from the
Monbukagakusho scholarship.
\par We acknowledge the use of the Legacy Archive for Microwave Background Data Analysis (LAMBDA). 
Support for LAMBDA is provided by the NASA Office of Space Science. 
\par All simulations, map operations and statistics were performed using software written by BL.

\end{acknowledgements}

\bibliography{current}  

\begin{thebibliography}{27}
\expandafter\ifx\csname natexlab\endcsname\relax\def\natexlab#1{#1}\fi

\bibitem[{{Aurich} {et~al.}(2004){Aurich}, {Lustig}, {Steiner}, \&
  {Then}}]{2004CQGra..21.4901A}
{Aurich}, R., {Lustig}, S., {Steiner}, F., \& {Then}, H. 2004, Classical and
  Quantum Gravity, 21, 4901, \eprint{astro-ph/0403597}

\bibitem[{{Bennett} {et~al.}(2003){Bennett}, {Hill}, {Hinshaw}, {Nolta},
  {Odegard}, {Page}, {Spergel}, {Weiland}, {Wright}, {Halpern}, {Jarosik},
  {Kogut}, {Limon}, {Meyer}, {Tucker}, \& {Wollack}}]{WMAPforegrounds}
{Bennett}, C.~L., {Hill}, R.~S., {Hinshaw}, G., {Nolta}, M.~R., {Odegard}, N.,
  {Page}, L., {Spergel}, D.~N., {Weiland}, J.~L., {Wright}, E.~L., {Halpern},
  M., {Jarosik}, N., {Kogut}, A., {Limon}, M., {Meyer}, S.~S., {Tucker}, G.~S.,
  \& {Wollack}, E. 2003, \apjs, 148, 97, \eprint{astro-ph/0302208}

\bibitem[{{Caillerie} {et~al.}(2007){Caillerie}, {Lachi{\`e}ze-Rey}, {Luminet},
  {Lehoucq}, {Riazuelo}, \& {Weeks}}]{2007AA...476..691C}
{Caillerie}, S., {Lachi{\`e}ze-Rey}, M., {Luminet}, J.-P., {Lehoucq}, R.,
  {Riazuelo}, A., \& {Weeks}, J. 2007, \aap, 476, 691, \eprint{0705.0217}

\bibitem[{{Cornish} {et~al.}(1998{\natexlab{a}}){Cornish}, {Spergel}, \&
  {Starkman}}]{Corn98a}
{Cornish}, N.~J., {Spergel}, D., \& {Starkman}, G. 1998{\natexlab{a}}, \prd,
  57, 5982

\bibitem[{{Cornish} {et~al.}(1998{\natexlab{b}}){Cornish}, {Spergel}, \&
  {Starkman}}]{Corn98b}
{Cornish}, N.~J., {Spergel}, D.~N., \& {Starkman}, G.~D. 1998{\natexlab{b}},
  Classical and Quantum Gravity, 15, 2657

\bibitem[{{Cornish} {et~al.}(2004){Cornish}, {Spergel}, {Starkman}, \&
  {Komatsu}}]{2004PhRvL..92t1302C}
{Cornish}, N.~J., {Spergel}, D.~N., {Starkman}, G.~D., \& {Komatsu}, E. 2004,
  Physical Review Letters, 92, 201302, \eprint{astro-ph/0310233}

\bibitem[{{de Oliveira-Costa} \& {Smoot}(1995)}]{deOliv95}
{de Oliveira-Costa}, A., \& {Smoot}, G.~F. 1995, \apj, 448, 477

\bibitem[{{Dineen} {et~al.}(2005){Dineen}, {Rocha}, \&
  {Coles}}]{2005MNRAS.358.1285D}
{Dineen}, P., {Rocha}, G., \& {Coles}, P. 2005, \mnras, 358, 1285,
  \eprint{astro-ph/0404356}

\bibitem[{{Gomero} \& {Rebou{\c c}as}(2003)}]{2003PhLA..311..319G}
{Gomero}, G.~I., \& {Rebou{\c c}as}, M.~J. 2003, Physics Letters A, 311, 319,
  \eprint{gr-qc/0202094}

\bibitem[{{G{\'o}rski} {et~al.}(2005){G{\'o}rski}, {Hivon}, {Banday},
  {Wandelt}, {Hansen}, {Reinecke}, \& {Bartelmann}}]{Healpix2005}
{G{\'o}rski}, K.~M., {Hivon}, E., {Banday}, A.~J., {Wandelt}, B.~D., {Hansen},
  F.~K., {Reinecke}, M., \& {Bartelmann}, M. 2005, \apj, 622, 759,
  \eprint{astro-ph/0409513}

\bibitem[{{Hinshaw} {et~al.}(2007){Hinshaw}, {Nolta}, {Bennett}, {Bean},
  {Dor{\'e}}, {Greason}, {Halpern}, {Hill}, {Jarosik}, {Kogut}, {Komatsu},
  {Limon}, {Odegard}, {Meyer}, {Page}, {Peiris}, {Spergel}, {Tucker}, {Verde},
  {Weiland}, {Wollack}, \& {Wright}}]{2007ApJS..170..288H}
{Hinshaw}, G., {Nolta}, M.~R., {Bennett}, C.~L., {Bean}, R., {Dor{\'e}}, O.,
  {Greason}, M.~R., {Halpern}, M., {Hill}, R.~S., {Jarosik}, N., {Kogut}, A.,
  {Komatsu}, E., {Limon}, M., {Odegard}, N., {Meyer}, S.~S., {Page}, L.,
  {Peiris}, H.~V., {Spergel}, D.~N., {Tucker}, G.~S., {Verde}, L., {Weiland},
  J.~L., {Wollack}, E., \& {Wright}, E.~L. 2007, \apjs, 170, 288,
  \eprint{astro-ph/0603451}

\bibitem[{{Hinshaw} {et~al.}(2003){Hinshaw}, {Spergel}, {Verde}, {Hill},
  {Meyer}, {Barnes}, {Bennett}, {Halpern}, {Jarosik}, {Kogut}, {Komatsu},
  {Limon}, {Page}, {Tucker}, {Weiland}, {Wollack}, \&
  {Wright}}]{2003ApJS..148..135H}
{Hinshaw}, G., {Spergel}, D.~N., {Verde}, L., {Hill}, R.~S., {Meyer}, S.~S.,
  {Barnes}, C., {Bennett}, C.~L., {Halpern}, M., {Jarosik}, N., {Kogut}, A.,
  {Komatsu}, E., {Limon}, M., {Page}, L., {Tucker}, G.~S., {Weiland}, J.~L.,
  {Wollack}, E., \& {Wright}, E.~L. 2003, \apjs, 148, 135,
  \eprint{astro-ph/0302217}

\bibitem[{{Inoue}(1999)}]{Inoue99}
{Inoue}, K.~T. 1999, Classical and Quantum Gravity, 16, 3071

\bibitem[{{Key} {et~al.}(2007){Key}, {Cornish}, {Spergel}, \&
  {Starkman}}]{2006astro.ph..4616S}
{Key}, J.~S., {Cornish}, N.~J., {Spergel}, D.~N., \& {Starkman}, G.~D. 2007,
  \prd, 75, 084034, \eprint{astro-ph/0604616}

\bibitem[{{Kunz} {et~al.}(2006){Kunz}, {Aghanim}, {Cayon}, {Forni}, {Riazuelo},
  \& {Uzan}}]{2006PhRvD..73b3511K}
{Kunz}, M., {Aghanim}, N., {Cayon}, L., {Forni}, O., {Riazuelo}, A., \& {Uzan},
  J.~P. 2006, \prd, 73, 023511, \eprint{astro-ph/0510164}

\bibitem[{{Kunz} {et~al.}(2008){Kunz}, {Aghanim}, {Riazuelo}, \&
  {Forni}}]{2008PhRvD..77b3525K}
{Kunz}, M., {Aghanim}, N., {Riazuelo}, A., \& {Forni}, O. 2008, \prd, 77,
  023525, \eprint{0704.3076}

\bibitem[{{Lehoucq} {et~al.}(1999){Lehoucq}, {Luminet}, \& {Uzan}}]{LLU98}
{Lehoucq}, R., {Luminet}, J.-P., \& {Uzan}, J.-P. 1999, \aap, 344, 735,
  \eprint{gr-qc/9604050}

\bibitem[{{Luminet} {et~al.}(2003){Luminet}, {Weeks}, {Riazuelo}, {Lehoucq}, \&
  {Uzan}}]{2003Natur.425..593L}
{Luminet}, J.-P., {Weeks}, J.~R., {Riazuelo}, A., {Lehoucq}, R., \& {Uzan},
  J.-P. 2003, \nat, 425, 593, \eprint{astro-ph/0310253}

\bibitem[{{Niarchou} \& {Jaffe}(2007)}]{2007PhRvL..99h1302N}
{Niarchou}, A., \& {Jaffe}, A. 2007, Physical Review Letters, 99, 081302

\bibitem[{{Niarchou} \& {Jaffe}(2006)}]{2006AIPC..848..774N}
{Niarchou}, A., \& {Jaffe}, A.~H. 2006, in American Institute of Physics
  Conference Series, Vol. 848, Recent Advances in Astronomy and Astrophysics,
  ed. N.~{Solomos}, 774--778

\bibitem[{{Phillips} \& {Kogut}(2006)}]{2006ApJ...645..820P}
{Phillips}, N.~G., \& {Kogut}, A. 2006, \apj, 645, 820,
  \eprint{astro-ph/0404400}

\bibitem[{{Riazuelo} {et~al.}(2004){Riazuelo}, {Uzan}, {Lehoucq}, \&
  {Weeks}}]{2004PhRvD..69j3514R}
{Riazuelo}, A., {Uzan}, J.-P., {Lehoucq}, R., \& {Weeks}, J. 2004, \prd, 69,
  103514, \eprint{astro-ph/0212223}

\bibitem[{{Roukema}(2000)}]{Roukema00-3}
{Roukema}, B.~F. 2000, \mnras, 312, 712, \eprint{astro-ph/9910272}

\bibitem[{{Roukema} {et~al.}(2004){Roukema}, {Lew}, {Cechowska}, {Marecki}, \&
  {Bajtlik}}]{RLCMB04}
{Roukema}, B.~F., {Lew}, B., {Cechowska}, M., {Marecki}, A., \& {Bajtlik}, S.
  2004, \aap, 423, 821, \eprint{astro-ph/0402608}

\bibitem[{{Sachs} \& {Wolfe}(1967)}]{SW67}
{Sachs}, R.~K., \& {Wolfe}, A.~M. 1967, \apj, 147, 73

\bibitem[{{Spergel} {et~al.}(2007){Spergel}, {Bean}, {Dor{\'e}}, {Nolta},
  {Bennett}, {Dunkley}, {Hinshaw}, {Jarosik}, {Komatsu}, {Page}, {Peiris},
  {Verde}, {Halpern}, {Hill}, {Kogut}, {Limon}, {Meyer}, {Odegard}, {Tucker},
  {Weiland}, {Wollack}, \& {Wright}}]{2007ApJS..170..377S}
{Spergel}, D.~N., {Bean}, R., {Dor{\'e}}, O., {Nolta}, M.~R., {Bennett}, C.~L.,
  {Dunkley}, J., {Hinshaw}, G., {Jarosik}, N., {Komatsu}, E., {Page}, L.,
  {Peiris}, H.~V., {Verde}, L., {Halpern}, M., {Hill}, R.~S., {Kogut}, A.,
  {Limon}, M., {Meyer}, S.~S., {Odegard}, N., {Tucker}, G.~S., {Weiland},
  J.~L., {Wollack}, E., \& {Wright}, E.~L. 2007, \apjs, 170, 377,
  \eprint{astro-ph/0603449}

\bibitem[{{Weeks}(2003)}]{2003MPLA...18.2099W}
{Weeks}, J.~R. 2003, Modern Physics Letters A, 18, 2099,
  \eprint{astro-ph/0212006}

\end{thebibliography}
\bibliographystyle{aa_hyperref}

\appendix

\section{Power spectrum, cut sky and smoothing length dependence on S correlations.}
\label{app:Sdependences}
\subsection{S dependence on the CMB power spectrum}
\par
As mentioned in Sect.~\ref{sec:data_and_sims} the $S$ correlation value depends not only on the particular alignment
of CMB fluctuations, but also on the underlying CMB \twopt correlation function (or its Legendre transform -- power spectrum),
as it is also a two point statistic. Therefore any discrepancies of the data from the assumed model will be \langed{statistically} 
imprinted 
onto the $S$ values evaluated from simulated maps (generated according to \langed{the} assumed model). This will 
lead to biases in \langed{estimates}
of the confidence level thresholds. Given that the nature of the large scale (mostly quadrupole and octupole)
anomalies of the WMAP data w.r.t. concordance LCDM model is unknown, we assumed \langed{a} model independent approach for 
generating simulations as described in Sect.~\ref{sec:statistics}. 
In this section \langed{we show} the impact of biasing of $S$ values due to large scale uncertainties in the assumed CMB power spectrum.

\par In order to quantify the impact of the underlying CMB power spectrum on the $S$ correlator values,
we perform the following exercise.\\
We create two INC simulations of the WMAP three year data, for two different power spectra, 
yet keeping exactly the same phase information in both cases.\\
One simulation is made using exactly the power spectrum reconstructed from the WMAP observations
as in Sect.~\ref{sec:data_and_sims} (i.e. neglecting cosmic variance effects). \\
The other simulation is made using a random realization of the best fit LCDM model \citep{2003ApJS..148..135H}
but with the same phase information as in the first simulation.
We Gaussian smooth them to common resolution with
beam of FWHM$=1^\circ$. This \langed{guarantees} that the differences in S values will only be due to
different underlying power spectra\footnote{The noise component is negligible at the considered smoothing scales of $1^\circ$}.
More specifically, since the LCDM model yields a good fit to the WMAP data for large $l$'s, the discrepancies will be only due to the
large scale anomalies.
\langed{Using $20$ random orientations of the fundamental dodecahedron, 
we then calculate the average of differences of
the $S$ correlator values between the two maps.}
That is, we calculate: $\langle \Delta S\rangle = \langle S_{LCDM}- S_{WMAP} \rangle$.
Hence, the positive $\langle \Delta S\rangle$ values show the excess correlations that one would additionally get if the LCDM model
was assumed, as compared to the ``exact'' realization of the reconstructed WMAP power spectrum (and \langed{ \begin{it}vice-versa\end{it} } for the negative values).

\par
\begin{figure}[!hbt]
\centering
\renewcommand{\figurename}{Fig}
\includegraphics[angle=-90,width=0.49\textwidth]{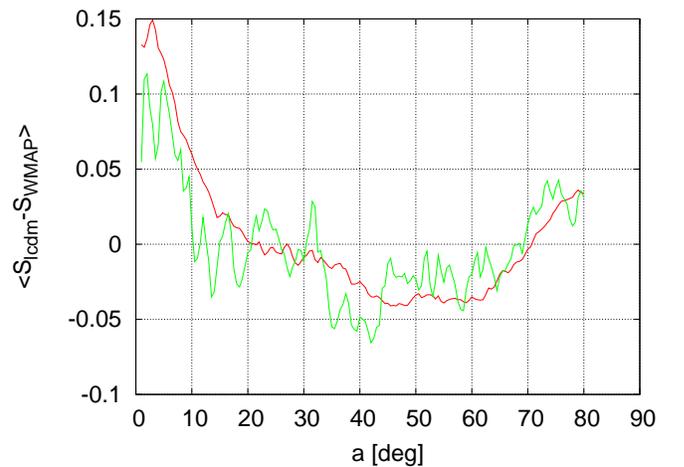}
\caption{Average difference $\langle \Delta S\rangle = \langle S_{LCDM}- S_{WMAP} \rangle$ correlator values, as a function
of circle size, statistically probing  different regions of the underlying power spectrum (red line).  With green line also
the difference relation is plotted for the orientation of the dodecahedron depicted in Fig.~\ref{fig:thedodec}.}
\label{fig:Sdiffav}
\end{figure}

 In Fig.~\ref{fig:Sdiffav} the $\langle \Delta S\rangle$ relation is shown with red line. 
Clearly, the strongest additional correlations appear for the smallest circles, i.e. for the largest angular separations
in the \twopt correlation function, and are as large as $\sim 0.15$. 
(These scales are additionally contaminated due to large smoothing scales as compared to the
circle size). While for the circle sizes of about $a\sim10^\circ$ the effect is small, it is obvious that at smaller circle
sizes the confidence level contours obtained from the simulations performed according to the LCDM model would be too conservative
than it is needed. Therefore, in this regard, given that the LCDM model is currently widely accepted, we consider our analysis to be very conservative.
This plot includes the dependence of $\langle \Delta S\rangle $ for larger circle sizes than those which we analyzed.
And it is easily seen that for scales of about $40^\circ$ the confidence level intervals would be underestimated. 

\par We therefore conclude that the simply connected space hypothesis would be consistent with the data
at yet even smaller CL, if the LCDM model \langed{were} assumed for the generation of simulations.

\subsection{S dependence on galactic sky cut}
\label{sec:S-gal-cut}
\par In this section we show the effects of different galactic sky cuts on the resulting amplitude of the correlation
signal. For this purpose we work with the first year WMAP ILC map, smoothed to a resolution of $2^\circ$ without mask.

\par
\begin{figure}[!hbt]
\centering
\renewcommand{\figurename}{Fig}
\includegraphics[angle=-90,width=0.49\textwidth]{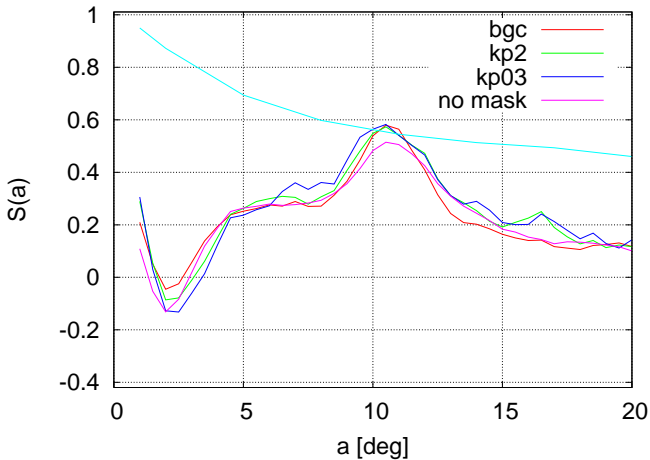}
\caption{Effects of different sky masks on the $S(a)$ correlation statistic. See text for mask abbreviations definition 
(Sect.~\protect\ref{sec:S-gal-cut}).
One sigma CL contour derived for the case of Kp2 is also plotted. }
\label{fig:cut-sky-dependence}
\end{figure}

 We compute the the correlation statistics for the orientation of the dodecahedron found in \cite{RLCMB04} and 
confirmed in this work (i.e. for the dodecahedron with the following face centers \thedodec and their opposites).
We then apply different sky \langed{masks} and show the $S(a)$ relations in Fig.~\ref{fig:cut-sky-dependence}.
We used three different sky masks: which we call ``bgc'', ``Kp2'', and ``kp03''.
The first of these is the one that was used in \cite{RLCMB04} (i.e. the galactic plane is masked for 
$|b| < 2^\circ$, and points within $20^\circ$ from the Galactic Center are masked). 
The second and third sky masks correspond to the Kp2 and the third year Kp0 sky masks \citep{WMAPforegrounds}.
\par Clearly, the fine details on the sky masking in case of ILC maps do not have large impact on the resulting $S$
correlation value, except for the ``no mask'' case where the correlation peak is lower (most likely due to some residual foreground 
contaminations of the Galactic plane).

\langed{We do not show the galactic sky cut dependence for the INC dataset.
This is because there are stronger foregrounds present than for the
ILC dataset, so that it is not straightforward to presmooth the map
without contaminating the ``clean'' regions of the sky by the Gaussian
tails of smoothing kernel.}

\subsection{S dependence on smoothing length}
In this section we present the impact of different smoothing lengths used during map smoothing process on the $S$
correlation values. We show the dependence using the dodecahedron orientation corresponding to the highest correlation
value ($\Smax$ value) at $a=12^\circ$ in Fig.~\ref{fig:ILC1INC3} (middle-right) which corresponds to the dodecahedron with faces 
\lbdod{49,51}{91,-2}{144,40}{256,64}{333,23}{28,-10} and their opposites. This closely matches the dodecahedron
depicted in Fig.~\ref{fig:thedodec}. We use \langed{the} WMAP three year INC map \langed{and the} Kp2 sky mask.
\par
\begin{figure}[!hbt]
\centering
\renewcommand{\figurename}{Fig}
\includegraphics[angle=-90,width=0.49\textwidth]{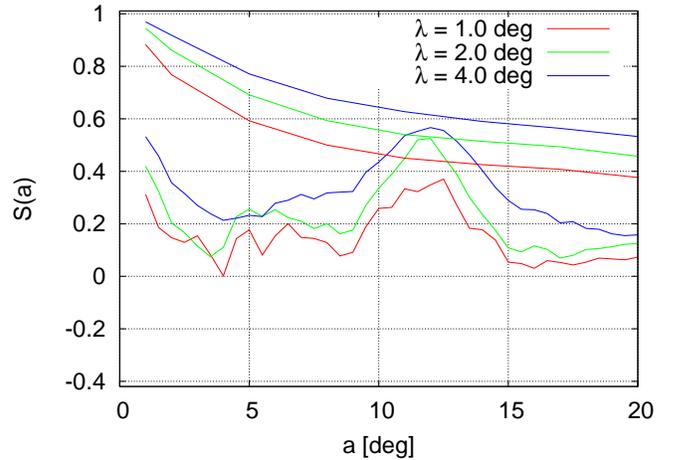}
\caption{Effect of different smoothing lengths on the $S(a)$ correlation statistic. We plot $S(a)$ dependence for the three consecutive smoothing lengths used in the 
analysis. Corresponding $1\sigma$ confidence thresholds for each smoothing scale are also shown. }
\label{fig:smooth-dependence}
\end{figure}

 In Fig.~\ref{fig:smooth-dependence} we \langed{show} 
the $S(a)$ values for three different smoothing lengths
used in the analysis. The characteristic trend towards higher correlation coefficients 
as smoothing length increases is clearly apparent, but the the 68\% CL contours increase by about the same amount. 

\end{document}